\documentclass{esub2acm}
\usepackage{pifont} 
\begin{document}
\title{A One-Time Pad based Cipher for Data Protection\\ in Distributed Environments}
\author{Igor Sobrado\\University of Oviedo}
\begin{abstract}
A \textit{one-time pad} (OTP) based cipher to insure both data protection
and integrity when mobile code arrives to a remote host is presented. Data
protection is required when a mobile agent could retrieve confidential
information that would be encrypted in untrusted nodes of the network; in
this case, information management could not rely on carrying an encryption
key. Data integrity is a prerequisite because mobile code must be
protected against malicious hosts that, by counterfeiting or removing
collected data, could cover information to the server that has sent the
agent. The algorithm described in this article seems to be simple enough,
so as to be easily implemented. This scheme is based on a non-interactive
protocol and allows a remote host to change its own data on-the-fly and,
at the same time, protecting information against handling by other hosts.
\end{abstract}
\begin{bottomstuff}

\begin{authinfo}
\affiliation{Facultad de Ciencias, Universidad de Oviedo}
\address{Avda. Calvo Sotelo 18, E-33007, Oviedo (Asturias), Spain; email:
\texttt{sobrado@acm.org}.}
\end{authinfo}

\permission
\end{bottomstuff}

\markboth{I. Sobrado}{A Cipher for Data Protection in Distributed Environments}
\maketitle

\section{Introduction}

Mobile agents are a promising technology that could improve performance
over the classical client/server model in distributed application
scenarios \cite{ismail:performance} and world-wide communications. It is
well known that agents offer important benefits in distributed
environments. Amongst other things, mobile agents reduce network load,
overcome network latency, encapsulate protocols, are executed
asynchronously and autonomously, adapt dynamically, are heterogeneous and
are fault-tolerant \cite{lange:reasons}. Mobile agents are also used to
search for distributed information \cite{menczer:textual}, in electronic
commerce \cite{morin:hypernews,ardissono:stores,ma:commerce}, network
management (service reconfiguration, mobility)
\cite{tennenhouse:networks}, control \cite{spoonhower:telephony},
monitorization, information streams automation, active networks
\cite{tennenhouse:active} and active documents. Looking at security,
mobile agents have a flaw. They are vulnerable to attacks from malicious
nodes and other agents in the network. A data protection scheme developed
to reduce security risks in distributed environments will be presented in
this work.

Mobile code based computing requires data management models that allow
information to be protected against both potential malicious \textit{peer
hosts} (remote hosts in the network) and hostile agents that could falsify
and destroy data collected by other agents or stored in nodes in the
\textit{information network}. Furthermore, peer hosts and mobile code
requires code area, execution thread and data part all to be protected
against both unauthorized disclosure and modification of information and
\textit{denial of service} (DoS) attacks
\cite{fong:linking,neuenhofen:marketplace}. In fact, comparable protection
requirements should be implemented on hosts and mobile code
\cite{vitek:computations}. Protection of hosts could be easily solved
using classical protection strategies like firewalls and unpermissive
access control policies; some examples are the use of \textit{access
control lists} (ACLs) and independent address spaces. However security of
agents require the development of new protection algorithms because
information will be managed in untrusted environments. It is known that
when mobile code arrives to a remote host it can be completely examined by
the host before running. At this moment, the remote host could counterfeit
or erase retrieved data to hide important information to the server that
has sent the mobile code. A classical approximation to solve this problem
is the use of \textit{cryptography}. In conventional cryptographic
algorithms a single key is used for both encryption and decryption. Such
systems, also known as \textit{symmetric}, require the key to be stored on
a safe place.  Obviously, symmetric cryptosystems are not suitable to
protect information collected by mobile code if the encryption key needs
to be conveyed with mobile code. \textit{Asymmetric ciphers} (public-key
cryptosystems) are a better choice to protect information against
malicious hosts, but are too slow and sometimes allow detectable patterns
in a message to survive the encryption process making the technology
vulnerable to cryptanalysis. To prevent this weakness, public-key ciphers
should hide these patterns by standard compression of the message before
encryption \cite{zimmermann:cryptography}.

\begin{table}
\centering
\caption{Operators Used in this Paper}
\label{operators}
\begin{tabular}{ll} \hline
Operator  & Description \\ \hline
$a \leftarrow b$ & assignation of value stored in $b$ to $a$ \\
$a = b$ & compares the values of $a$ and $b$ \\
$f \oplus g$ & concatenation of bit fields $f$ and $g$ \\
$f\,\&\,g$ & bitwise \textit{and} operation between bit fields $f$ and $g$ \\
$f\,\texttt{xor}\,g$ & bitwise \textit{exclusive-or} operation between bit fields $f$ and $g$ \\
$f \ll n$ & an $n$-bit rotation (not displacement) to the left of the bit field $f$ \\
$f \gg n$ & an $n$-bit rotation (not displacement) to the right of the bit field $f$ \\
$array[i]$ & denotes the $i$-th element of the vector $array[1 \ldots n]$ \\
$array[i]:i=1,2,\ldots,n$ & denotes the entire vector $array[1 \ldots n]$ \\ \hline
\end{tabular}
\end{table}

\begin{figure}
\centering
\epsfig{file=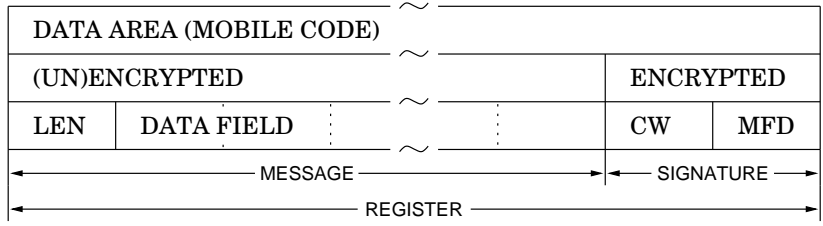}
\caption{Each register in the mobile code data area contains a message
field that is composed by the message length (LEN) and data (DATA) fields.
The signature includes the codeword (CW) and the message field digest
(MFD) which are those that will be used to authenticate the message.}
\label{register}
\end{figure}

\begin{table}
\centering
\caption{Classification of Keys Used}
\label{keys}
\begin{tabular}{lll} \hline
Key                   & Length               & Notes                           \\ \hline
Digital signature key & 128-bit              & Could be transmitted using an untrusted \\
                      &                      & communication channel.\\\\
Encryption key        & 128-bit greater      & Requires a secure communication \\
                      & than data field size & channel.                        \\ \hline
\end{tabular}
\end{table}

For security reasons public-keys must be certified before using.
Information networks based on public-key ciphers requires to have
\textit{certification authorities} (CAs) that will be used to authenticate
public-keys provided by the nodes in the network. These CAs could be hosts
in the information network or external hosts shared between two or more
networks. It is easy to see that encryption keys must be certified.
Suppose as an example that our organization has an information network
where mobile code retrieves data using a public-key cipher to authenticate
all information delivered. One host wants to protect the provided data
using its own public, $Pub_{H_{i}}$, and private, $Pri_{H_{i}}$, keys,
where $Pub$ and $Pri$ denotes public and private keys respectively and
$H_{i}$ labels the $i$-th host in the mobile agent route. If public-keys
are not certified, any host in the information network could simulate a
pair of private and public-keys for the data provider using its own pair
of private ($Pri_{H_{i}}^{false}$) and public ($Pub_{H_{i}}^{false}$) keys
to digitally sign messages supplying
\begin{displaymath}
 sgn_{i,j}^{false} \leftarrow f_{Pri_{H_{i}}^{false}}(msg_{false})
\end{displaymath}
instead of
\begin{displaymath}
 sgn_{i,j} \leftarrow f_{Pri_{H_{i}}}(msg)
\end{displaymath}
as the $j$-th message for the $i$-th host in the agent route. Here $sgn$
stands for a signed message and $f$ for an encryption function. Table
\ref{operators} presents a description of the operators used in several
sections of this article. In this case, the server that has sent mobile
code has no way to determine the host that really signed the data. The
same problem occurs when encryption is used. Now,
\begin{displaymath}
 cpr_{i,j}^{false} \leftarrow
 f_{Pub_{S}}\Big[f_{Pri_{H_{i}}^{false}}(msg_{false})\Big]
\end{displaymath}
could be used by a malicious host replacing the real ciphertext ($cpr$)  
provided by the $i$-th host in the route of the agent:
\begin{displaymath}
 cpr_{i,j} \leftarrow f_{Pub_{S}}\Big[f_{Pri_{H_{i}}}(msg)\Big],
\end{displaymath}
where $Pub_{S}$ denotes the public-key of the agent server. Another
advantage related with CAs is that the public-key of a CA could be
used to authenticate and/or decrypt all messages certified by that CA. As
a consequence, only one public-key is required to manage the information
provided by all the hosts associated with a particular CA.

The algorithm described in this work does not rely on a public-key
cryptosystem and only requires to establish a connection with the remote
information network hosts to acquire a copy of the \textit{one-time keys}
that were used to authenticate or encrypt information collected by mobile
code. In Figure \ref{register} the structure of the registers stored in
the mobile agents data area is shown. Table \ref{keys} presents a
classification of keys used by our protection algorithm, the size of these
keys and some additional notes related with keys management. This
connection does not rely on an interactive protocol between the agent and
the server that has sent it. A communication channel to share used keys
will solely be established when mobile code returns to the agent server.
Only connections established between remote hosts in the network and the
\textit{agent server} in order to share one-time encryption keys will
require a secure communication channel. When exclusively data integrity
must be assured, once mobile code has arrived to the server that has sent
it, the data management model proposed in this paper allows one-time keys
to be sent, over untrusted channels, to the server generating the mobile
agent. This is because these keys will not be used again. In any case,
these keys will not be released before the agent returns. Most mobile
agent based applications only requires the protection of a part of the
mobile code data area. In particular, our information protection scheme
admits digital signature and encryption of provided data by peer hosts.
The agent data area can evolve dynamically as a consequence of visiting
servers where information is collected. The algorithm that is going to be
described in this work allows data protection in a way that prevents
unauthorized modification or disclosure of information but grants a host
to change its own information\footnote{For security reasons a peer host
could not change information provided by other nodes in the information
network.}.

The way our protection scheme manages data encryption and digital
signature is one without carrying cryptographic keys and without requiring
interaction between the server which has sent the mobile code and the
proper agent. It is implemented such that only authorized hosts (the host
that provides information and the server that has sent the mobile agent)
could change the retrieved data with simultaneous protection against
``brainwash''. To avoid data erasing, activity of agents will be logged by
external hosts in the network.

\section{Sharing keys between peer hosts and the agent server}

\begin{figure}
\centering
\epsfig{file=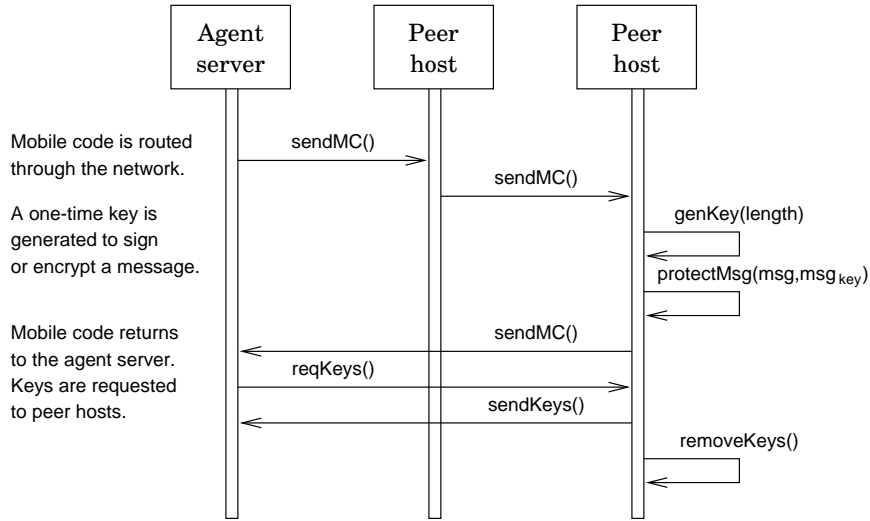}
\caption{Digital signature and encryption of data will require one-time
keys that will be requested to the peer hosts when mobile code returns to
the agent server. Each remote host must check its own generated keys to
assure that were not used to protect another previously signed message
stored in the mobile code data area.}
\label{communication}
\end{figure}

A security model for mobile code should conceive protocols requiring
minimal interaction between the server that has sent the mobile code and
the agent itself \cite{sander:towards}. The proposed data management model
does not rely on an interactive protocol between the mobile code and the
server that has sent it and that would like to go off-line. A secure
connection between remote hosts in the network and the agent server, for
example using the \textit{transport layer security} (TLS)
protocol\footnote{TLS is the latest revision of the \textit{secure socket
layer} (SSL) protocol.} \cite{dierks:tls,lawrence:tls}, will only be
required to share encryption keys. Even without using such a protocol, it
is possible to implement secure communication channels over untrusted
networks with different algorithms \cite{abadi:channel}. If we only need
to digitally sign data and not to encrypt it, information can be sent to
the server as plain-text when mobile agent returns; in this case, keys
will not be compromised because the server will have a copy of data that
could not be counterfeited by peer hosts at this moment. These keys will
not be used again to digitally sign other registers in the future. Figure
\ref{communication} depicts the communication established between the
agent server and two peer hosts to route the agent through the network,
store and protect information in the data area of an agent and retrieve
the keys used. Table \ref{functions} lists a description of the functions
used in these algorithms.

\begin{table}
\centering
\caption{Description of Functions Used in the Algorithms}
\label{functions}
\begin{tabular}{ll} \hline
Function & Description \\ \hline
\texttt{checkMsg}$(msg,msg_{key})$ & authentication of message $msg$ using key $msg_{key}$ \\
\texttt{genKey}$(length)$ & makes a valid (unique) one-time key of size $length$ \\
\texttt{length}$(a)$ & returns the size of $a$ \\
\texttt{protectMsg}$(msg, msg_{key})$ & protects the message $msg$ using the key $msg_{key}$ \\
\texttt{rand}$(n)$ & creates an $n$-bit length random field \\
\texttt{removeKeys}$()$ & erases the generated keys from peer host data area \\
\texttt{reqKeys}$(hostid)$ & requests the generated keys applied by the peer host $hostid$ \\
\texttt{sendKeys}$(serverid)$ & sends the generated keys to the agent server $serverid$ \\
\texttt{sendMC}($hostid$) & sends the agent to the host $hostid$ \\ \hline
\end{tabular}
\end{table}

\begin{figure}
\centering
\epsfig{file=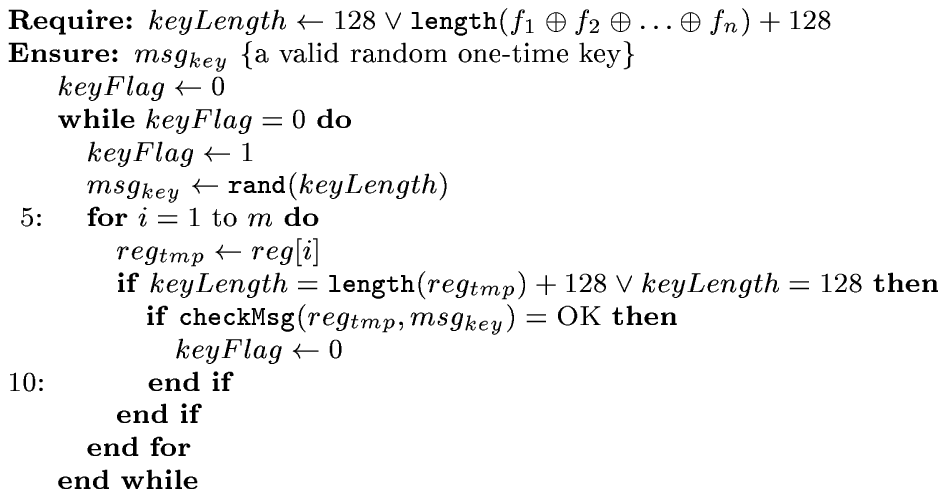}
\caption{\texttt{genKey}$()$: Random one-time keys generation algorithm.}
\label{genKey}
\end{figure}

\begin{figure}
\centering
\epsfig{file=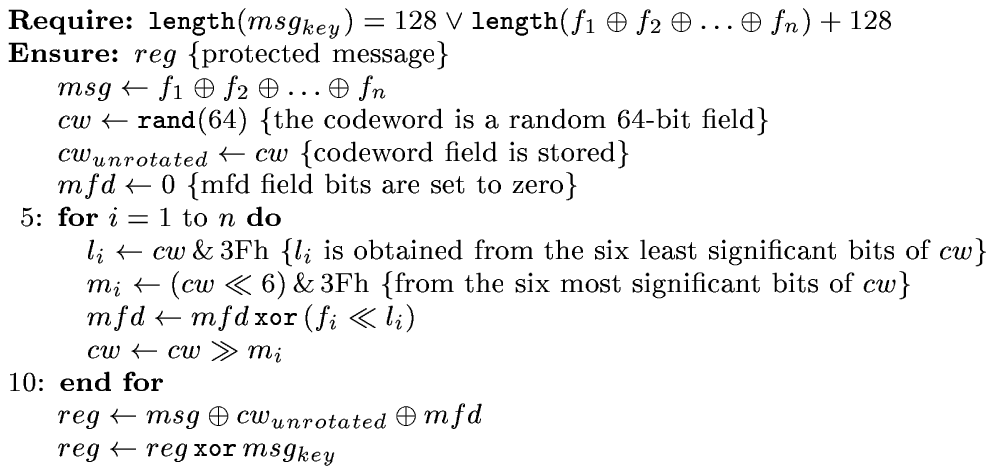}
\caption{\texttt{protectMsg}$()$: Digital signature and data encryption
algorithm.}
\label{protectMsg}
\end{figure}

Now, we are going to explain the propagation of keys. Each host in the
network that wants to provide some information must create a random
one-time key that will be used to digitally sign or encrypt just one
message using the function \texttt{genKey}$()$. In Figure \ref{genKey}
appears the random one-time keys generator proposed to create valid keys
to protect new registers for a mobile agent. The one-time key created by
the host will be applied to the message using the function
\texttt{protectMsg}$()$. Figure \ref{protectMsg} provides a description of
the algorithm used to protect data provided by peer hosts to the agent.
The function \texttt{genKey}$()$ will assure that this key could not be
applied to other messages in the mobile code data area before accepting
it. This restriction ensures that messages signed or encrypted by remote
hosts will be easily identified. These keys will be shared with the agent
server through a secure channel when data encryption is required. When the
agent is traveling in the network, keys must not be provided to other
hosts, in particular to the server that has sent it, because information
is vulnerable to attack or damage in untrusted nodes. The protocol
described in this article allows to establish a link between the agent
server and remote hosts in the network using the function
\texttt{reqKeys}$()$ to get copies of the applied keys when the agent
returns to the server that has dropped it. Whenever an agent server
requests the used keys to a remote host, after sending them to the server,
they must be removed by the generating host by using the function
\texttt{removeKeys}$()$.

\begin{figure}
\centering
\epsfig{file=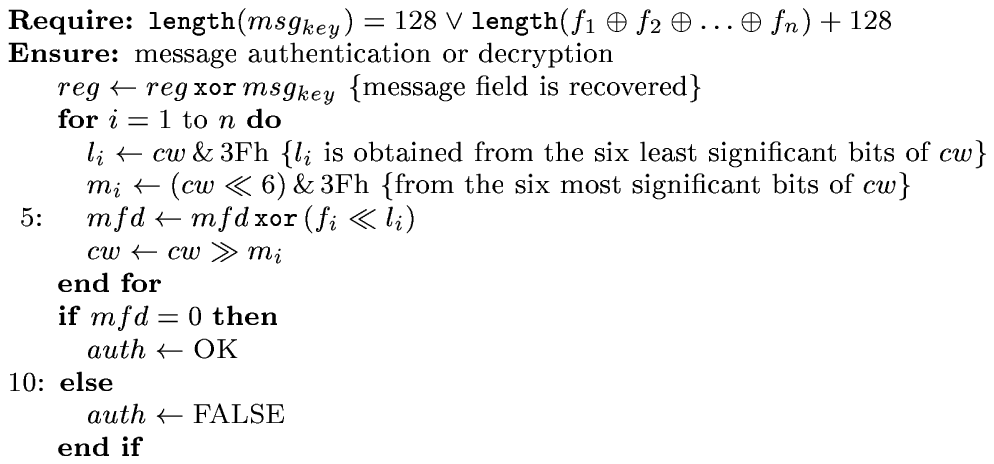}
\caption{\texttt{checkMsg}$()$: Authentication and decryption of data
algorithm.}
\label{checkMsg}
\end{figure}

If we forecast that mobile agents may travel through more than one
information network in the same trip, we must assure that keys will be
used only one time. It is highly unlikely, but possible, to sign or
encrypt two messages using the same key. We could think that mobile code
could verify this fact, but only if it does not imply to carry a copy of
the used keys in the mobile code data area. It is possible to check keys
generated by each peer host by trying to match it with every register
carried by the mobile code. That a random one-time key matchs a register
does not mean that we have obtained a key used by other host as it will be
shown below, but to assure security this key must be discarded. Figure
\ref{checkMsg} shows an algorithm that could be used to authenticate and
decrypt the registers stored in the mobile agent data area using the keys
created by the remote hosts. Upon the return of the agent, these keys will
be delivered to the server that owns it using the function
\texttt{sendKeys}$()$.

To prevent attacks trying to discover the keys at use, previously used
cryptographic keys will not be applied again. Instead of using the old key
a new one will be generated to sign or encrypt a message. As commented
above, each peer host should check that its own generated cryptographic
keys could not be applied to other registers. The check will be performed
by trying to apply these keys on the registers stored in the mobile agent
internal table. This condition assures that the host that signed the data
will be identified without problems by the server that generated the
agent.

\section{Retrieving and processing information}

At this moment we have a way to share encryption keys between remote hosts
in the information network and the agent server. As shown above, each
register provided by a peer host will be digitally signed or encrypted
using a different one-time key. These keys will be sent to the agent
server after the return of the agent. On this section we will show how to
digitally sign and encrypt information provided by remote hosts and how to
authenticate information retrieved using these shared keys.

\subsection{Retrieving information from peer hosts}

\begin{figure}
\centering
\epsfig{file=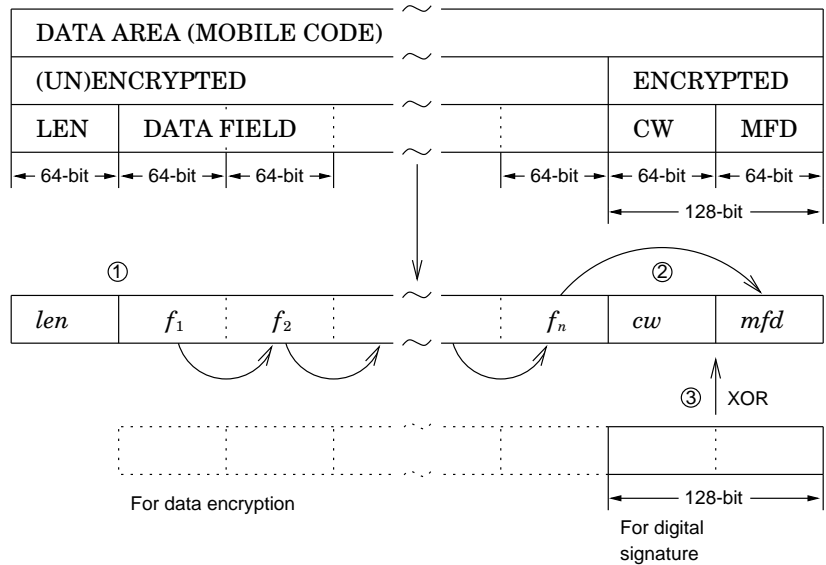}
\caption{Digital signature and encryption of data on remote hosts is based
on splitting the data field on equal-sized blocks (64-bit blocks). These
blocks, denoted as $f_{i}$ where $i = 1, 2, \ldots, n$, will be processed
in \ding{192} using the codeword to obtain the MFD field that will
authenticate the message itself, in \ding{193}. After that, the signature
must be covered using an exclusive-or bitwise operator. Whenever
encryption is required the data field will be covered with an encryption
key of the same size than the field itself that should be added to the
128-bit digital signature key used to protect the signature field of the
register. This is shown in \ding{194}.}
\label{process-1}
\end{figure}

In the following subsection, we are going to describe a detailed
information management in peer hosts, showing how to digitally sign and
encrypt data provided by remote hosts. A scheme to protect data against
``a mobile code brainwash'' based on logging the agent route in external
servers and use one-time keys will be examined too. Figure \ref{process-1}
is an overview of the process of digital signature and encryption of data.

\subsubsection{Digital signature and encryption of data}

Remote hosts should protect data provided by encrypting or, at least,
digitally signing its own information fields before releasing the mobile
agent. As shown above, they could digitally sign its own data with a
128-bit key or encrypt it using a cryptographic 128-bits key greater than
the data field size (as shown in Table \ref{keys}). The length field (LEN)
will store the data field length. This field is needed to provide a most
flexible communication protocol supporting messages of arbitrary size. It
is also needed because we have no way to define an end-of-message code for
our communication protocol if encryption is a requeriment.

\begin{itemize}
\item\textit{Digital signature.} Digital signature of information will
protect data against handling by other hosts in a safe way but allowing
reading. Digital signature of information does not permit counterfeited
information to be carried by mobile code and, as will be shown below, to
remove information provided by peer hosts. This is because digitally
signed fields can only be validated with the encryption keys requested to
the peer hosts by the agent server. It is possible because the algorithm
shown in this article uses a random 64-bit block that must be generated
for each message digitally signed by a remote host. This random block is
never sent as plain-text. This random field will be called
\textit{codeword} (CW) along this work. Only a host that have a copy of
the used key intended to cover the codeword and the message field digest
could easily authenticate information provided by remote hosts and
identify simultaneously the host that digitally signed it. A host that
does not have a copy of either the encryption key or the codeword
(therefore it could not obtain the message field digest that depends on
the data field and the codeword itself) will have $2^{64}$ possible
combinations that could be valid to digitally sign the message. We have no
way to know which key has been used to protect a message.
\item\textit{Data encryption.} Information could be easily changed in a
way that does not permit to recover blocks of the message and/or the
encryption key applied. If a server provides predictable data this fact
could not be used by a malicious host to change the message contents. The
message field digest will be obtained in the same manner as in the process
of digital signature but now the key used to protect the register covers
the data, codeword and message digest fields. This key is nothing but the
\textit{encryption key} in this article and obviously will need to be sent
to the agent server by means of a secure channel. The agent server could
apply these encryption keys provided by peer hosts to the encrypted
registers to recover the original message that will be checked in the same
way as the digital signature.
\end{itemize}

Let us suppose a mobile agent returns to a peer host and needs to change
some information previously posed in the agent data area. Trying to
decrypt the messages, this host can look for its own registers using the
encryption keys generated to protect the messages. This condition should
be checked with the message field digest. To prevent data erasing, remote
hosts must generate a new random key to sign the message discarding the
key currently used. In other case, a brute force attack against both the
old and the new messages would be possible making the protection algorithm
vulnerable. In the case a peer host needed to remove a register it could
look for it in the same way, remove it from the mobile agent data area and
erase the key used to protect the register itself so that it will not be
sent to the agent server in the future.

A message field digest will protect each register. The algorithm proposed
to generate and hide the message field digest is easy to implement in any
programming language and is fast, allowing a peer host to process big
volumes of information quickly. Let us take a look to Figure
\ref{process-1}. The data field must be splitted in 64-bit blocks. The
number of blocks that are needed to split a message is stored in the
message length field and will be used to determine the message length even
if encryption is requested. The field digest is set to zero and then the
next algorithm must be applied to each 64-bit block:
\begin{enumerate}
\item As shown in Figure \ref{process-2} the $i$-th 64-bit block must be
rotated $l_{i}$ bits to the left in \ding{192}, where $l_{i}$ is obtained
from the six least significant bits of the codeword field. The rotated
field will be stored in the message field digest using an exclusive-or
bitwise operator.

\begin{figure}
\centering
\epsfig{file=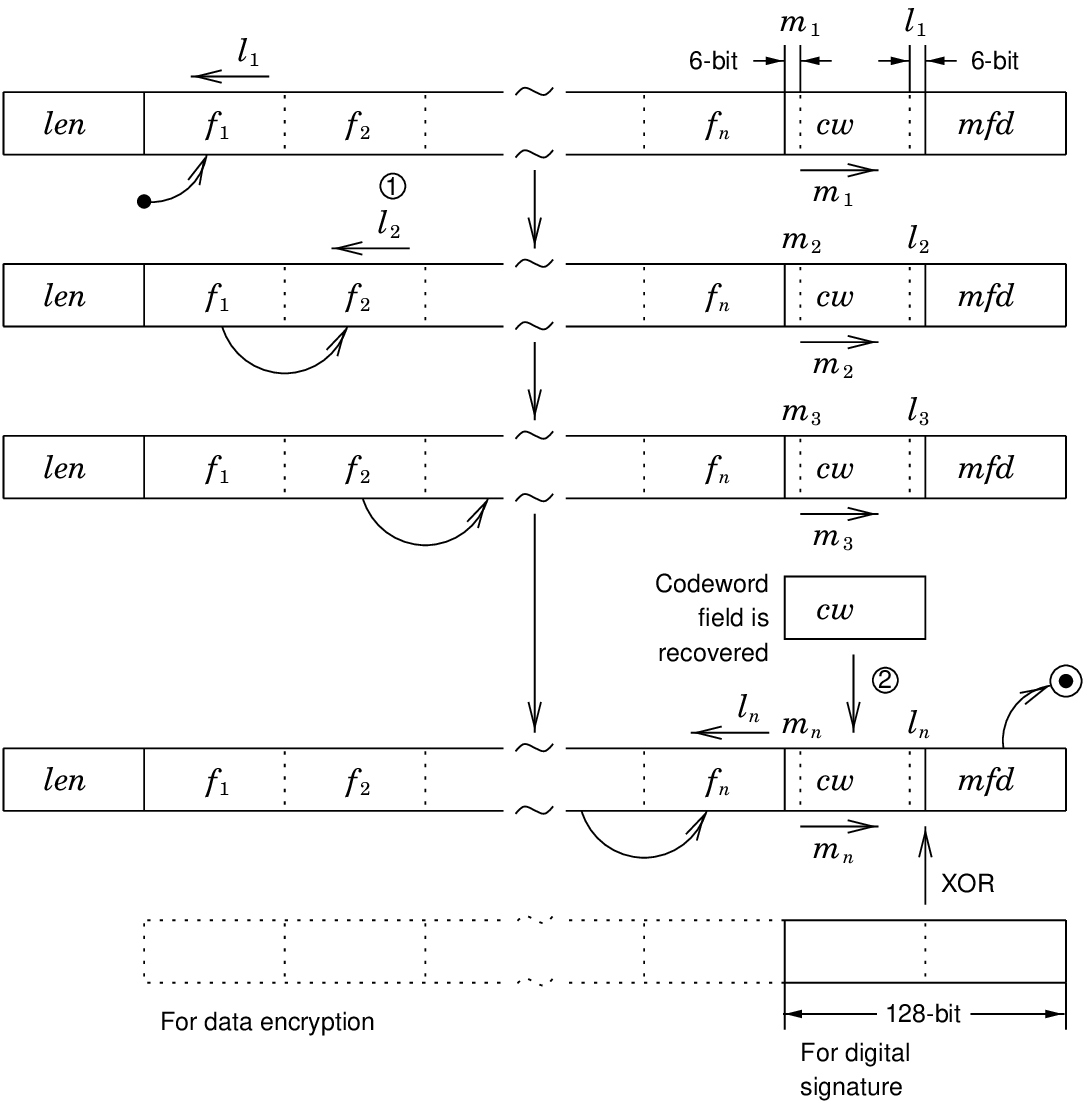}
\caption{Detailed description of digital signature and encryption
processes for a message. In this figure, $l_{i}$ and $m_{i}$, where $i =
1, 2, \ldots, n$, are respectively the values of the six less and most
significant bits of the codeword field on each rotation. Encryption
requires to overwrite the $f_{i}$ fields with the 64-bit fields rotated to
the left, as shown in \ding{192}, and to apply a 128-bit key greater than
the data field size (obtained adding dotted fields to the signature) to
protect information. Codeword field itself is rotated $m_{i}$ bits to the
right and recovered when process ends (in \ding{193}).}
\label{process-2}
\end{figure}

\item The codeword field itself is rotated $m_{i}$ bits to the right,
where $m_{i}$ is provided by the six most significant bits of the
codeword. This operation over the codeword field will assure that the
rotation of the next 64-bit field, the $(i+1)$-th 64-bit block, could not
be found.
\end{enumerate}

After obtaining the field digest, the codeword field must be overwritten
with the unrotated codeword, as shown in \ding{193} making easier to check
register integrity using the same routine for both encryption (or digital
signature) and decryption (or authentication of information). At last, the
codeword and the field digest will be protected with the random one-time
key (a 128-bit key for digital signature or a key 128-bit greater than
data field length for encryption) using an exclusive-or bitwise operator
hiding its contents as shown in \ding{194}. Figure \ref{process-2} shows
how fields will be rotated to generate the message digest and how this
field is protected against reading by unauthorized hosts. The digital
signature and encryption algorithm presented in this article is really
quick and effective with respect to computational requirements.

\subsubsection{Preventing a mobile code's brainwash}

\begin{figure}
\centering
\epsfig{file=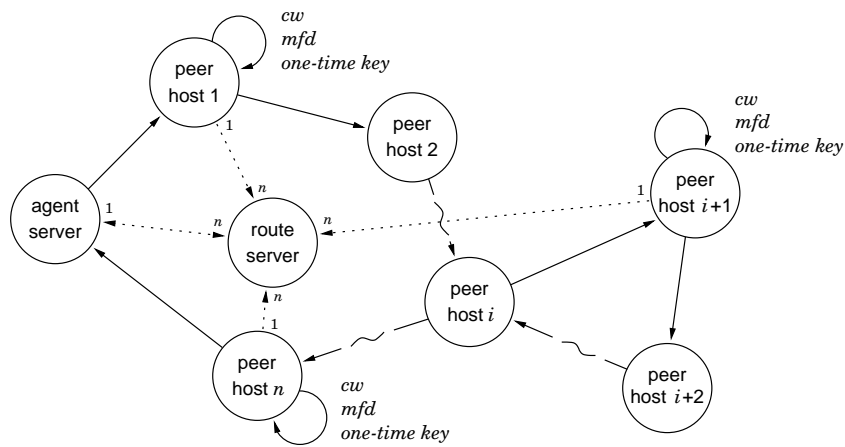}
\caption{One or more \textit{route servers} will provide a way to log the
mobile agent route to avoid data covering. If a route is not registered by
an external server, a malicious host (in our example the $i$-th peer host
in the route of the agent) might overwrite the data area of the mobile
agent with a bitcopy of an old data area when the agent returns then
hidding information delivered by hosts $i+1, i+2, \ldots$}
\label{route}
\end{figure}

Information provided by peer hosts must be protected against both
counterfeit and erasing. The former could be avoided using digital
signature or data encryption techniques as described above. The latter
requires a way to log agent activities in the network. Obviously, this
information could not be conveyed with the mobile agent because we cannot
protect it against unauthorized modification. Should a mobile agent return
to a malicious host, both data and log areas could be overwritten with a
bitcopy of old data and log areas carried by the agent covering
information provided since it was released for the first time by the
hostile host. To workaround this problem, we propose to store this
information in one or more remote \textit{route servers} (RSs). Figure
\ref{route} shows how RSs could be used to log the routes followed by
agents. This information could be sent to these servers over untrusted
communication channels without risk. These servers will provide the route
followed by each agent to the agent servers upon requested.

In our propousal, a host could delete its own information. As described
above, in the case a peer host needed to remove some information
previously provided to a mobile agent it would be able to erase the
message field that must be eliminated and the one-time key used to protect
the message itself simultaneously. This procedure will assure that the old
one-time key will not be sent to the agent server in the future. If the
agent server receives a key that could not be applied to a message stored
in the mobile agent data area, the information provided by that agent
could not be authenticated and then the ported data should be discarded.

\subsection{Authentication and decryption of data in the agent server}

When a mobile agent returns to the server that has sent it, information
coming from remote hosts must be authenticated and, if it was protected
against reading by other hosts, decrypted. This server will request to the
RSs the agent route information and, after that, the keys used by the peer
hosts. The agent server must try to apply all the cryptographic keys
provided by the peer hosts to the registers collected by the mobile agent
in order to try to identify the hosts that have signed the registers. As
we have shown above, these cryptographic keys were generated in peer hosts
and were not provided to other hosts in the network. As a consequence,
they are supposed not to be compromised (we are not talking about
malicious people administrating these hosts). When the agent arrives to
the server that has dropped it, keys could be requested over untrusted
communication channels\footnote{Obviously, only if data encryption was not
mandatory.} since the server that has sent the mobile agent already have a
copy of the data provided by peer hosts and these keys will not be
re-used. Only the right key will match the message field digest and will
decrypt the information if needed. This event can be easily checked using
the algorithm depicted in Figure \ref{checkMsg}. This analysis must be
applied to all the registers found on the data area.

The scheme described in this article allows a mobile agent to return to a
remote host an unlimited number of times even if it is a malicious one. A
security related restriction imposed by the algorithm described is that a
host can only change its own registers. It is easy to see that in any
other case it is not possible to assure data integrity.

\section{Security of the communication protocol}

Mobile agents and peer hosts require to be protected against attacks from
other agents and hosts in the network. In this section we will study the
security of the communication protocol against brute force attacks and
spoofing techniques intending to supplant a host identity.

\subsection{Protection against off-line guessing}

All strong password mechanisms proposed in the literature employ
public-key techniques. The possibility of develop a symmetric cipher safe
to brute force attacks has been posed in \cite{halevi:protocols}. Our
protection strategy is immune to off-line guessing. If encryption is
requested, security of the protocol depends on the security of the
communication channel used to share the symmetric one-time encryption keys
and the security of the remote hosts that store the keys.

\newtheorem{proposition}{Proposition}

\begin{proposition}A cryptosystem whose encryption key has the same length
as the message to protect is invulnerable against off-line guessing
attacks if we have a way to generate a perfect random one-time encryption
key.\end{proposition}

\begin{proof}Suppose we have an $n$-bit length message and a perfect (non
predictable) random encryption key with the same length. It is easy to see
that we can establish a bijective relation between the bits of the message
and the bits of the key. Let us suppose that the $i$-th bit of the
encryption key could change the state of the $i$-th bit of the message; as
a result, there are no statistical methods that any potential attacker
could hope to use to infer detectable patterns in the message and that
could allow the prediction of the final state of the bits in the message
from the ciphertext, what is a simple consequence of Information
Theory.\end{proof}

It is worth to remark that unpredictable encryption keys are needed. A
predictable (or at least partially predictable) encryption key will allow
statistical attacks against the ciphertext trying to discover parts of the
information provided. Good random generators have been proposed in the
bibliography \cite{press:recipes}. Next proposition assures that registers
are off-line guessing resistant.

\begin{proposition}For any given ciphertext $c[i]:i=1,2,\ldots,n$ we can
found an encryption key $b[j]:j=1,2,\ldots,n$ that allows to hide an
arbitrary message of the same length $a[k]:k=1,2,\ldots,n$ using the
cipher described in this work.\end{proposition}

\begin{proof}We have an $n$-bit length ciphertext $c[i]:i=1,2,\ldots,n$
obtained using an exclusive-or bitwise operator as described in subsection
3.1.1 (here $c[i] = a[i]\,{\rm xor}\,b[i]:i=1,2,\ldots,n$ where $a[i]$
denotes the $i$-th bit of the message and $b[i]$ labels the $i$-th bit of
the encryption key). It is possible to hide an arbitrary message $a[i]:i =
1, 2,\ldots, n$ if we define the key as:
\begin{displaymath} b[i] =
\left\{ \begin{array}{ll}
        0 & \mbox{if $a[i] = c[i]$} \\
        1 & \mbox{otherwise}
       \end{array}
       \right.
\end{displaymath}
where $i = 1, 2, \ldots, n$.\end{proof}

It is easy to see that off-line guessing could not be applied against
digital signatures and encrypted data fields because we cannot determine,
using brute force attacks, what key has been applied to digitally sign the
message.

\begin{proposition}For any given message we can obtain $2^{64}$ possible
digital signatures. By using brute force attacks based techniques, we have
no way to know what key has been applied.\end{proposition}

\begin{proof}Suppose we have a ciphertext $c[i]:i=1, 2, \ldots, 128$
provided with the message as digital signature. These ciphertext hides a
codeword (a random bitmap) $cw[i]:i = 1, 2, \ldots, 64$ and a message
field digest $mfd[i]:i = 1, 2, \ldots, 64$ that depends on the codeword
itself. It is easy to see that we have $2^{64}$ possible codewords and a
message field digest for each codeword field. Applying proposition 2 we
can find a key for each pair $(cw, mfd)$ that allows us to obtain the
ciphertext $c$.\end{proof}

As a consequence, we conclude that off-line guessing based attacks against
data protected using the algorithm described are not possible when digital
signature of data is applied. If encryption of data is required, security
depends on both classes of hosts (peer hosts and the agent server) and the
communication channel used to provide copies of the encryption keys
managed by the agent server.

\subsection{IP-spoofing attacks}

The current IP protocol technology (IPv4) does not allow to eliminate
IP-spoofed packets in the network. We are currently working in an
IPv6-based protection scheme that will offer a better solution to this
problem and will be presented elsewhere. To install filtering routers has
been proposed as the best protection practice in \cite{cert:ca-96.21} but
there is not a generally accepted solution to this threat yet. However for
the seek of completeness a strategy based on filtering routers could be
implemented as follows:
\begin{itemize}
\item\textit{input filters.} An input filter is a filtering router that
will restrict the input to the external interface of the route server.
This filter must block each package that has a source address belonging to
the internal network but comming from outside eliminating in this way the
possibility of superseding the identity of a internal host in the network.
\item\textit{output filters.} An output filter is a filtering router that
will refuse all packages coming from inside and with an external source
address avoiding then IP-spoofing attacks that could be originated from
any host of the subnet.
\end{itemize}
Obviously filtering routers will work only when the network where hosts
that provide information are placed is physically isolated from the
external network. We should not assume that the peer hosts where agents
will collect information are all in the same physical subnet. Neither can
we assume that all remote hosts are trusted. In fact, we cannot protect
hosts against spoofing techniques using filtering routers because the
hosts receiving the mobile agent could be spreaded over the whole network.
It is the mobility of the agent what breaks down the difference between
internal and external. Internet Protocol version 6 includes aditional
security features useful to protect mobile agents. It is then natural to
implement in this context good anti-spoofing techniques based on
authenticating the IP headers.

\section{Security considerations}

Security risks must be considered on the design of information networks
based on mobile code. As noted by \shortciteANP{neuenhofen:marketplace},
mobile agents are extremely vulnerable to all kinds of attacks from
potentially malicious peer hosts and agents once it leaves the agent
server. Weaknesses related with remote hosts and code protection (for
example, host protection against unauthorized accesses and DoS attacks,
host authentication and code protection against modification) will not be
treated here because are out of the scope of this work and have been
extensively considered in the bibliography \cite{hohl:mess,sander:hosts}.
In this section well known security risks related with data protection in
mobile agents will be studied.

\subsection{Security weaknesses in mobile agents environments}

Security weaknesses related with mobile computing has been extensively
considered in the bibliography. As mentioned above, one of the main
problems here is that of agents which will be executed in untrusted nodes.
These hosts will have full access to code and data areas of the agent (and
to the execution thread in mobile code too). Security risks related with
classical computing environments like off-line guessing attacks against
the ciphertext are also present. Some of these weaknesses can be itemized
as:

\textit{Attacks by hosts in the route of an agent.} Some protection
schemes proposed in the bibliography \cite{yee:sanctuary} rely on carrying
keys that will be used to obtain \textit{partial result authentication
codes} (PRACs). After using them, these keys will be destroyed (simple
MAC\footnote{MAC stands for \textit{message authentication code}.}-based
PRACs) or changed by means of \textit{one-way functions} (MAC-based PRACs
with one-way functions). In his work, Bennet S. Yee explains that the
former could be easily attacked copying un-removed keys making vulnerable
any host $s_{j}:i < j < n$, where $s_{i}$ is the malicious host and
$s_{n}$ is the last peer host in the mobile code route. The latter is
based on calculating, when mobile code is going from host $s_{i}$ to host
$s_{i+1}$, the key $k_{i+1} = f(k_{i})$ where $f$ is a one-way function. A
malicious host $s_{i}$ could get the key $k_{i}$ and obtain
$k_{i+1}=f(k_{i}), \,k_{i+2}=f(k_{i+1}),\ldots,\,k_{n}=f(k_{n-1})$ turning
vulnerable the hosts $s_{i+1},\,s_{i+2},\ldots,\,s_{n}$. Publicly
verifiable PRACs allow an agent itself to check the partial results
obtained allowing computations that depend on previous partial results
trying to detect any integrity violation of those results. This can be
achieved without arriving at the server that has sent it.

\textit{Off-line guessing.} As described above, symmetric ciphers require
the keys to be stored in a safe place. In this case, keys must not be
safely sent over untrusted channels. Asymmetric ciphers allow public-keys
to be sent over untrusted communication channels but exposing detectable
patterns in a message that survive the encryption process making these
public-key based ciphers vulnerable to cryptanalysis whenever the message
is not compressed before encryption. Both symmetric and asymmetric ciphers
are vulnerable to off-line guessing (brute force attacks against the
ciphertext).

\textit{Data erasing.} All protection schemes applied to mobile agents
allow information to be protected against counterfeit. These algorithms
could be used to avoid data handling by non-authorized hosts in the
network in the sense that information could not be modified, but sometimes
they allow information to be partially or fully removed from the data area
of the agent.

\subsection{How our system solves these weaknesses}

Our goal is to protect the registers in the data area for both agents and
mobile code against counterfeit and erasing in a way that will be
transparent to final users. The protection of the code area and the
execution thread, if present, is beyond the scope of this article.
Exhaustive works dwelling on security concepts in mobile agents have been
developed during the last years
\cite{hohl:mess,sander:towards,vitek:computations}. These works are mainly
devoted to protect the code area of mobile agents but not the data area
that, as we mentioned above, will evolve dinamically when mobile agents
travel over the network.

\textit{Attacks by hosts in the route of an agent.} The protection scheme
proposed in this work does not require the keys to be sent with the mobile
code. Instead of this, each host will generate a random one-time key for
digital signature or encryption of data. These keys will not be carried in
the mobile code data area and will not be provided to other hosts.
Consequently, an agent can return to a previously visited peer host,
including malicious hosts, safely. A remote host can not change data
provided by any other host without invalidate the register carried by the
mobile code. As we have extensively described, a mobile agent data area
``brainwash'' is not possible because the routes followed by agents will
be registered by the RSs.

\textit{Off-line guessing.} Some problems related with off-line guessing
could not be avoided. If encryption is required, one-time keys used to
protect data must be transmitted over secure channels; as a result, data
security depends on channel security. Digital signature does not require
keys to be transmitted using secure channels. As we already noted, keys
will not be stored as a part of the agent data state and will then not be
delivered to the agent server before the agent returns. We have already 
shown that keys could not be guessed using brute force attacks.

\textit{Data erasing.} A possible workaround could be to store information
about the hosts visited in remote RSs instead of in the mobile agent data
area. Each host that wants to provide information to the agent could
protect itself sending its own network address to the RSs proposed by the
agent server. Each host in the network could extract this information from
the agent code area. Redundant information could eventually be sent to the
agent server when keys are requested and could also be used to rebuild the
route followed by the agent.

\section{Other advantages of the algorithm}

The protection algorithms applied to agents must satisfy the restrictions
imposed by mobile agent based systems. A minimal interaction between the
agent server and the agent itself is one of this restrictions. Each host
that provides information must be able to modify or remove its own
registers without invalidate the agents.

\textit{Non-interactive protocol.} The proposed data protection algorithm
is based on a non-interactive protocol between the agent server and the
mobile code. As a consequence, the server that has sent an agent have the
possibility of going off-line. This is one of the main goals of a mobile
code based system. So, the use of a non-interactive protocol reduces
bandwidth requirements.

\textit{Keys do not need to be carried together with the mobile agent.}
Our algorithm does not require keys to be carried with agents, but allows
peer hosts to generate its own random one-time keys in a way that assures
that they are completely new keys. It is important to notice that this
does not mean that a brute force attack against ciphertext trying to
discover used keys is possible. One might think that the freedom of a host
to generate lots of keys could be a way of guessing and checking actually
used keys but anyone of these would be one over $2^{64}$ possible
choices. The only possibility the host has is that of being very lucky
picking with a single trial the correct one among such a big set. The root
of this indeterminacy is the fact that underneath of the 128-bit signature
there is a random field we have called codeword and that affects to the
message digest.

\textit{Information could be changed.} This algorithm is such that
information provided by peer hosts could be changed when needed by the
host that has generated it. It is interesting to compare our proposal with
that of Tschudin's \cite{tschudin:security} that consists in appropiately
linking an information with any other one coming from other hosts to
accomplish that a malicious host will not be able to remove the signed
data. This is because it can not forge the signatures of the other hosts.
There is here an advantage in the sense that, in principle, RSs are not
needed. But there is a clear inconvenience because in the best case we can
finally find ourselves with a message containing a lot of obsolete
information that could not be removed by its owner.

\section{Related work}

Strong foundation is a requirement for future work in the topic of mobile
agents \cite{kendall:patterns}. To design semantics and type-safety
languages for agents in untrusted networks \cite{riely:trust} and
supporting permissions languages for specifying distributed processes in
dynamically evolving networks, as the languages derived from the
$\pi$-calculus \cite{riely:typed} are important to protect hosts against
malicious code. \shortciteANP{spoonhower:telephony} have shown that agents
could be used for collaborative applications reducing network bandwidth
requeriments. \shortciteANP{sander:hosts} have proposed a way to obtain
code privacy using non-interactive evaluation of encrypted functions
(EEF). \shortciteANP{hohl:mess} has proposed the possibility of use
algorithms to ``mess up'' code.

We are currently developing a public-key based cipher that will solve this
problem using standard cryptographic tools but at a higher computational
cost. The main advantage of this algorithm will be that it would avoid the
need to wait for a network link to share a cryptographic key because all
hosts in the information network and the server that has sent the mobile
code would get copies of public-keys from a key server. These keys could
be certified by a CA. We are working on a key propagation algorithm that
will make easier to share public-keys in distributed environments.

\section{Conclusions}

Mobile and distributed computing requires new security schemes that do
not depend on carrying cryptographic keys. During the past, information
security was based on symmetric ciphers because cryptographic keys were
stored in servers accessible only through firewalls and control access
procedures. However, data management models are changing quickly. Agent
based computing is fundamental in networking environments and, as shown
before, cannot be based on classical protection schemes. Our encryption
method, as other cryptographic algorithms, obeys the following 
requeriments:
\begin{itemize}
\item\textit{It is not needed to hide the algorithm} nor the digital
signature keys, only the encryption keys must be protected against
reading by unauthorized hosts;
\item\textit{The encryption process must destroy statistical parameters}
and the structure and predictable patterns in the language;
\item\textit{An error in the transmission should not destroy the rest of
the information provided} but, obviously, will invalidate the message
digest.
\end{itemize}

The data management model shown in this article allows data signature and
encryption in a way that only authorized hosts (the server that has sent
mobile code and the host that provides information) could modify and
certificate information retrieved then, at the same time, protecting data
against
``mobile code brainwash'' by malicious hosts. The cryptographic algorithm
described is faster and easier to implement on mobile code
environments than public-key based ciphers that have a higher
computational cost. Other important property of this algorithm is that it
does not allow brute force attacks against the ciphertext.

\begin{acks}
The author would like to thank Dar\'{\i}o \'Alvarez and Jes\'us Arturo
P\'erez for introduce him in the distributed computing world. Thanks go to
M. A. R. Osorio, Diego Rodr\'{\i}guez and Jos\'e Luis Ru\'{\i}z for many
helpful comments and reviewing the draft of the article. The author
greatly appreciate the Department of Mathematics of the University of
Oviedo for donating him forever a SPARC workstation from Sun Microsystems
that has been used in the current research.
\end{acks}

\bibliographystyle{esub2acm}

\end{document}